# Positive and negative pressure effects on the magnetic ordering and the Kondo effect in the compound, $Ce_2RhSi_3$

T. Nakano,[*] Kausik Sengupta,[a] Sudhindra Rayaprol,[a] M. Hedo, Y. Uwatoko and E.V. Sampathkumaran[a,$]

*The Institute for Solid State Physics, The University of Tokyo, 5-1-5 Kashiwanoha, Kashiwa, Chiba 277 8581, Japan*
[a]*Tata Institute of Fundamental Research, Homi Bhabha Road, Mumbai-400005, India*

**Abstract**

The competition between magnetic ordering and the Kondo effect in $Ce_2RhSi_3$, ordering antiferromagnetically at 7 K, is investigated by the measurements of magnetization, heat capacity and electrical resistivity on the solid solutions, $Ce_{2-x}La_xRhSi_3$, $Ce_{2-y}Y_yRhSi_3$, and $Ce_2RhSi_{3-z}Ge_z$, as well as by high pressure studies on this compound. The trends in the Kondo and Néel temperature variations among these alloys are compared to infer the roles of unit-cell volume and electronic structure changes. On the basis of the results, we infer that this compound lies at the peak of Doniach's magnetic phase diagram. The high pressure electrical resistivity data indicate that the quantum critical point for this compound is in the vicinity of 4 GPa.

I. Introduction

The Ce-based intermetallic compounds are at the centre stage of the field of 'strongly correlated electron behavior'. In this respect, the study of the competition between the Kondo effect and magnetic ordering continues to be a fruitful direction to pursue, as most of the discoveries of various phenomena (e.g., heavy-fermion superconductivity, non-Fermi liquid behavior) in this field are essentially a consequence of this direction of research. Therefore, it is of interest to carry out such investigations in new Ce compounds. In this article, we address this competition phenomena in the Kondo lattice, $Ce_2RhSi_3$, crystallizing in a $AlB_2$-derived hexagonal structure [1, 2] as very little work has been done on this compound, though it was discovered few decades ago. In this compound, Ce is trivalent and was established to order anti-ferromagnetically near 7 K by bulk measurements; the neutron diffraction studies [3, 4] reveal that the magnetic moment of 1.3 $\mu_B$ at 1.3 K lies parallel to the b-axis of an orthorhombic magnetic unit cell. There is some ambiguity in the exact space-group this class of ternary rare-earth compounds belong to, but it appears that the *P6_3/mmc* space group is the most probable one [5]. Despite the fact that there are two sites for Ce (called *2(b)* and *6(h)*) in this crystal structure, the measurements till to date could reveal only one magnetic transition, as indicated by a well-defined peak in the plot of magnetic susceptibility ($\chi$) versus temperature (T) [3]. The plot of heat-capacity (C) divided by T versus $T^2$ is linear in the paramagnetic state below 20 K and the linear term was found to be about 100 mJ/mol $K^2$; it is not clear whether it represents true electronic term, as possible interference from short range magnetic order and Schottky effects due to crystal-field effects in this temperature range. For our purpose, we have investigated the influence of positive and negative chemical pressure by gradual substitution of Ce by Y and La and Ge for Si on the magnetic and transport behavior of this compound. The main difference between La/Y doping on the one hand and Ge doping on the other is that in the former cases there is a dilution of the Kondo lattice, whereas in the latter, the Ce sub-lattice is undisturbed. The initial results on Y and La substituted samples have been reported in a conference proceedings [6]. In addition, we have carried out magnetization (M) and electrical resistivity ($\rho$) studies under external pressure up to 8 GPa.

II. Experimental details

The samples, $Ce_{2-x}La_xRhSi_3$, $Ce_{2-y}Y_yRhSi_3$, and $Ce_2RhSi_{3-z}Ge_z$, (x= y= 0.0, 0.3, 0.5, 1.0, 1.5, 1.7 and 2.0; z= 0.2, 0.4, 0.6, 0.8 and 1.0), were prepared by arc melting stoichiometric amounts of constituent elements, followed by homogenization in evacuated sealed tubes at 800 C for 5 days. The samples were characterized by x-ray diffraction (Cu $K_\alpha$) and the lattice constants are listed in Table 1. Typical x-ray diffraction patterns are shown in figure 1. We are not able to synthesize single phase compositions for higher values of *z* in the Ge substituted alloys. Dc magnetization measurements (T= 1.8-300 K) were carried out employing a commercial superconducting quantum interference device (SQUID) (Quantum Design, USA). The heat-capacity data (1.8 – 15 K) were collected with a Physical Property Measurements System (PPMS) (Quantum Design, USA). The $\rho$ measurements were performed (1.8-300 K) for all these alloys by a four-probe method employing a conducting silver paint for making electrical contacts. For the parent Ce compound, we have also performed $\rho$ measurements under external pressure employing a cubic anvil high pressure system up to 8 GPa down to 1.8

K using a mixture of Flourinert FC-70 and FC-77 as a pressure transmitting medium; measurements were extended with a hybrid cylinder (NiCrAl, CuBe) pressure cell down to 50 mK up to 3.2 GPa in a dilution refrigerator using Daphne oil as a pressure transmitting medium. The magnetization measurements (2-300 K) under pressure (Cu-Be cell) were also performed up to 1.5 GPa in the SQUID magnetometer.

### III. Results and discussion
### A. Magnetic susceptibility

In figure 2, we show dc $\chi$ behavior obtained in a magnetic field (H) of 5 kOe in the form of $\chi^{-1}$ versus T for all the alloys. As reported earlier [3], for x= 0.0, the plot is linear above 100 K and at lower temperatures there is a deviation from linearity presumably due to crystal-field effects and/or the Kondo effect. The value of the effective moment ($\mu_{eff}$= 2.46$\mu_B$/Ce) obtained from the linear region is typical of trivalent ions; the paramagnetic Curie temperature ($\theta_p$) is about -65 K. Since the magnetic ordering sets in at much lower temperatures (below 7 K), clearly such a large magnitude of $\theta_p$ (with a negative sign) implies the existence of the Kondo effect. The values of $\theta_p$ are given in table 1. In the case of Kondo lattices, naively speaking, one could expect a contribution from indirect exchange interaction as well to $\theta_p$. However, in these alloys, this contribution to $\theta_p$ is negligible, noting that the values of $\theta_p$ even for heavy rare-earth members are much smaller [1] (which means that the de Gennes scaled $\theta_p$ value for the Ce case is smaller than 1 K). Therefore, it is safe to assume the well-established fact [7] that this parameter is directly related to single-ion Kondo temperature ($T_K$) for these Kondo alloys. As La(Y) concentration is increased, the negative sign of $\theta_p$ is retained for all Ce containing compositions, but the magnitude decreases (increases). In the case of La series, the value of $\theta_p$ is about -30 K for x= 1.5, whereas the corresponding composition in Y series is characterized by a value of -180 K. From these values, it is clear that La(Y) substitution depresses (enhances) $T_K$. In the case of Ge substituted alloys, $\theta_p$ shows a decreasing trend with increasing Ge concentration. Since, in the Ge series, the Ce sublattice is not disturbed unlike in La series, one can confidently state that the unit-cell volume (V) change induced by an expansion of the lattice bears an observable effect on the depression of $T_K$ [8]. A careful comparison of the unit-cell volume values (see table 1) of La and Ge substituted alloys, however, indicates that the lattice expansion by the substitution at the Si site by Ge is more effective in reducing $T_K$, as an increase of V by about 3 Å$^3$ in the case of former decreases $\theta_p$ by few Kelvin only, whereas in the latter, the value falls by as much as about 15 K. This comparison reveals that there are possible electronic structure variations following Ge substitution for Si which play a dominant role to depress $T_K$. With respect to Ce valence behavior, the lattice compression caused by Y substitution does not cause any change in the valence of Ce, as the value of $\mu_{eff}$ per Ce is practically the same as that of the parent compound.

In order to infer the trends in the variation of $T_N$, we have performed $\chi$ measurements in the presence of a field of 5 kOe after cooling the sample to 1.8 K in zero field, that is, for the zero-field-cooled (ZFC) condition of the specimen (see figure 3). As mentioned earlier, there is a well-defined peak due to magnetic ordering near 7 K for the parent Ce compound in $\chi(T)$. This peak shifts to lower temperatures with increasing La concentration; in the case of Y series, magnetic ordering manifests itself as a tendency of $\chi$ to flatten. The temperature at which these features appear are taken as a measure of $T_N$. It is clear that both La and Y substitutions cause similar effects on $T_N$. For compositions more than 1.0, the magnetic transition shifts below 2 K, as indicated by the characteristic features described above. For the Ge substituted alloys, the features due to magnetic ordering appear at the nearly same temperature; that is, the transition temperature is not strongly dependent on $z$. It should however be noted that $T_K$ as indicated by $\theta_p$ decreases by about 25% as $z$ is increased from 0.0 to 1.0. This implies that the peak region in the Doniach's phase diagram [9] is reasonably broad.

**B.     Heat-capacity**

We infer the trends in $T_N$ variation on the basis of the features in C(T) plots as well. The results of C measurements are shown in figure 4 for relevant compositions. It is distinctly clear that, for compositions less than 1.5 in the La and Y substituted alloys, there are features attributable to magnetic ordering in the form of a peak in C(T). Though it is very difficult to precisely determine the value of $T_N$ due to broadened C(T) peaks, one can draw a qualitative conclusion by comparing the C(T) curves that there is a gradual depression of $T_N$ with increasing La or Y. The point to be noted is that both La and Y substitutions are almost equally effective in depressing $T_N$ for a given composition supporting the conclusions from figure 3. [Following Ref. 10 for such broadened C(T) curves, we arbitrarily define the middle point at the rising curve in C(T) with decreasing temperature as $T_N$, marked by an arrow in figure 4 for instance for x= 0.0, for the purpose of table 1]. However, for a given volume change, say by about 7 Å$^3$, the $T_N$ value is much smaller for La series compared to that for Y series (compare the C(T) curves for x= 1.0 and y= 0.3). This may imply that La is more effective in depressing $T_N$, possibly emphasizing the role of electronic structure as well on the magnetic properties. In order to infer the role of unit-cell volume on $T_N$ in the alloys, we have obtained the value of $T_N$ divided by 2-x (or 2-y) (considering validity of indirect exchange interaction in metals), and compared it with that of parent compound. It is found that such scaled-values do not exceed that of parent compound. This finding suggests that the parent compound lies close to the peak of Doniach's magnetic phase diagram [9]. Consistent with this finding, for Ge substituted samples, the features due to magnetic ordering occur nearly at the same temperature range for all compositions; however, a careful look at the peak temperatures indicates that there could be a marginal upward shift (by about 1 K) of $T_N$ as $z$ is increased from 0 to 1, as though the parent compound can be placed marginally to the right of the peak in the Doniach's diagram.

**C.     Electrical resistivity**

We now present the results of $\rho$ measurements, mainly to support above conclusion. We attribute apparent large value of residual resistivity ratio to strong Kondo scattering effects and we do not think that it is due to disorder. For the parent compound, $\rho$ is nearly constant above 150 K (see figure 5) and there is a gradual fall below 150 K followed by an upturn below 15 K till $T_N$. This kind of feature in $\rho$(T) in the paramagnetic state is typical of an interplay between the Kondo effect and the crystal-field effect. We have also extracted the 4f contribution ($\rho_{4f}$) to $\rho$ and for this purpose we have employed the $\rho$(T) values of the La analogue. Two logarithmic regions can be found in the plot of $\rho_{4f}$, supporting the existence of above-mentioned interplay. The ratio of the high to low temperature slopes of the $\rho_{4f}$ plot is about 0.12, which is very close to that expected [11] for the crystal-field-split ground state. Figures 6 and 7 show the data for La and Y alloys, normalized to respective 300 K values. As Ce is replaced by La/Y, the drop due to magnetic order gets gradually depressed towards lower temperatures in agreement with the $\chi(T)$ and $C(T)$ data and the upturn below 20 K persists (see figure 7). For x and y greater than 1.0, the upturn is only observed without any feature due to magnetic ordering in the temperature range of investigation. Thus, the $\rho(T)$ curves exhibit the features attributable to magnetically-ordered Kondo lattice to Kondo-impurity transformation as Ce sublattice is diluted. There is a qualitative difference in the $\rho(T)$ curves of La and Y substituted alloys. There is a distinct broad maximum in the $\rho(T)$ plot around 150 K for y>0.3 for the Y series, which is absent for the La series. Thus an interplay between the Kondo effect and the crystal-field effects is visible even in the raw data in the case of Y series. Finally, the temperatures at which the drop in $\rho$ occurs due to the onset of magnetic ordering are in qualitative agreement with those inferred from the $C(T)$ plots. In the case of Ge substituted alloys also (Figure 7), $\rho(T)$ plots establish that $T_N$ values are essentially unchanged for all compositions investigated.

**D.     High pressure electrical resistivity and magnetization behavior of $Ce_2RhSi_3$**

The results of high pressure $\rho$ experiments employing cubic anvil cell are shown in figures 8 (2-300 K) and 9 (2-40 K) for the parent Ce compound. The shapes of the curves are gradually modified, particularly at low temperatures (<30 K) with increasing pressure. The double-peaked structure due to the interplay between the Kondo and crystal-field effects vanishes for $P \geq 2$ GPa establishing that the positive pressure gradually increases 4f hybridization strength thereby broadening crystal-field levels. For P = 6GPa and 8 GPa, we observe $T^2$ dependence of $\rho$ at low temperatures without any evidence due to magnetic ordering (above 1.8 K), whereas the variation is linear with T below about 10 K for 4 GPa as though, at this pressure, there is a tendency for non-Fermi liquid behavior. It thus appears that there is a crossover from magnetic ordering to nonmagnetism around 4 GPa; that is, quantum critical point (QCP) may occur around this pressure. We have extended high pressure studies to mK range in a dilution refrigerator to track how $T_N$ varies in the low pressure regime and the measurements could be performed up to 3.2 GPa only with this pressure cell. The data thus obtained are shown in figure 10. It is clear that an initial application of pressure, say 0.83 GPa, depresses the temperature at which $\rho(T)$ curve changes the slope due to magnetic ordering (5.8 K). With a further increase of pressure, the feature due to magnetic ordering gets smoothened and hence it is rather difficult to find $T_N$ at higher pressures. Nevertheless, from a small change of slope, we infer that the temperature marked by arrows is a good estimate of $T_N$. Thus, for P= 3.2 GPa, $T_N$ is around 2.7 K. Though there is some ambiguity on estimating $T_N$ in this manner, there is no doubt that $T_N$ decreases with increasing pressure as inferred from figure 10, in broad agreement with the findings based on Y substituted series. We have also confirmed this trend by performing high pressure dc magnetic susceptibility studies (in a magnetic field of 10 kOe) at low temperatures, the results of which are shown in figure 11. It is clear that the features due to magnetic ordering are well-defined even under external pressure and the peak in $\chi(T)$ shifts down to a lower temperature with increasing pressure, with its position at about 5.5 K for 1.5 GPa, in qualitative agreement with the trend inferred from the $\rho$ data. A profile of $T_N$ as a function of P inferred from all these data is shown in the inset of figure 11. (Note: The shape of the curve below $T_N$ for P= 0.8 GPa is different from those in other curves and the origin of the difference is not clear to us at present).

Finally, we would like to explore whether there are any differences in the magnetic structure induced by positive and negative pressure without disrupting Ce sub-lattice. For this purpose, we compare the isothermal magnetization behavior at 1.8 K for Ge substituted alloys and pressurized $Ce_2RhSi_3$ in figure 12. There is a curvature [3] in M-H plot at 1.8 K near 35 kOe for the parent compound due to spin reorientation, which is a signature of antiferromagnetism in zero-field. We observe that this curvature diminishes for P= 1.1 GPa and is totally absent for 1.5 GPa. On the other hand, for Ge substituted samples, the curvature in M-H plots (at 1.8 K) is retained or more pronounced with increasing z. This difference may indicate that there are subtle differences in the changes in the magnetic structure with negative and positive chemical pressure.

**IV      Summary**

The positive and negative pressure effects on the Kondo and magnetic ordering behavior of a compound, $Ce_2RhSi_3$, have been investigated. The results establish that this compound lies near the peak in the Doniach's magnetic phase diagram.  Features attributable to 'magnetically-ordered Kondo lattice' to 'Kondo impurity' transformation are seen in the electrical transport data of La and Y substituted alloys. It appears that the lattice expansion caused by Ge substitution at the Si site is much more effective in depressing the Kondo temperature than that by La substitution at the Ce site as indicated by the trends in paramagnetic Curie temperature values, though the data for the latter series indicate that  the electronic structure changes following Ge substitution play an observable role. However, for a given volume change, lattice expansion caused by La substitution is more effective in depressing $T_N$ than that caused by lattice compression by Y substitution, implying that possible changes  in the electronic structure, apart from volume change, also play a role on the changes in the magnetic properties.  High pressure results reveal that the shape of $\rho(T)$ curve undergoes profound qualitative changes with external pressure, and particularly near 4 GPa, quantum critical point possibly exists. While we do not find evidence for superconductivity up to 3.2 GPa down to 50 mK, it is of interest to perform low temperature studies in the vicinity of QCP (say  at 4 GPa) to characterize non-Fermi liquid characteristics as well as to search for superconductivity at QCP.  Finally, studies on single crystals of $Ce_2RhSi_3$ are highly desirable, but unfortunately our attempts to prepare the single crystals are unsuccessful.

Acknowledgements:
We would like to thank Kartik K Iyer for his valuable experimental help during the course of this investigation.  One of us (EVS) would like to acknowledge the an invitation by the Institute for Solid State Physics to perform high pressure work.

References:
$Corresponding author. E-mail address: sampath@tifr.res.in
*Present address: *Waseda University, Tokyo 169-8555, Japan*
1.    B. Chevalier et al., Solid State Commun. **49**, 753 (1984).
2.    R.E. Gladyshevskii et al., J. Alloys Comp. **189**, 221 (1992).

3. I. Das and E.V. Sampathkumaran, J. Magn. Magn. Mater. **137,** L239 (1994).
4. J. Leceijewicz, N. Stüsser, A. Szytula, and A. Zygmunt, J. Magn. Magn. Mater. **147**, 45 (1995).
5. W. Bazela, E. Wawrzynska, B. Penc, N. Stüsser, A. Szytula, and A. Zygmunt, J. Alloys and Compounds, **360,** 76 (2003); R. –D. Hoffmann and R. Pöttgen, Z. Kristallogr. **216**, 127 (2001).
6. Kausik Sengupta, S. Rayaprol and E.V. Sampathkumaran, Proceedings of Department of Atomic Energy (India) Symposium held in 2003, vol 46, 817 (2003); arXiv.org cond-mat/0309701.
7. G. Gruener and A. Zawadowski, Progress in Low Temperature Physics, Vol. VII B, edited by D.R. Brewer (North-Holland, Amsterdam).
8. J.W. Allen and R.M. Martin, Phys. Rev. Lett. 49, 1106 (1982).
9. S. Doniach, Physica B 91, 231 (1977).
10. M. Bouvier, P. Leuthuilier, and D. Schmitt, Phys. Rev. B 43, 13137 (1991)
11. B. Cornut and B. Coqblin, Phys. Rev. B **5**, 4541 (1972).

**Table 1:** The lattice constants ($a, c, \pm 0.004$ Å), unit-cell volume ($V$), paramagnetic Curie temperature ($\theta_p \pm 2K$), effective magnetic moment ($\mu_B \pm 0.05 \mu_B$) and Néel temperature ($T_N$) for the alloys, $Ce_{2-x}La_xRhSi_3$, $Ce_{2-y}Y_yRhSi_3$, and $Ce_2RhSi_{3-z}Ge_z$.

| x, y, z | | $a$ (Å) | $c$ (Å) | $V$ (Å$^3$) | $\theta_p$ (K) | $\mu_{eff}$($\mu_B$/Ce) | $T_N$ (K) |
|---|---|---|---|---|---|---|---|
| | 0 | 8.240 | 8.444 | 496.49 | -65 | 2.46 | 7.0 |
| x= | 0.3 | 8.240 | 8.476 | 498.81 | -63 | 2.47 | 5.8 |
| | 0.5 | 8.243 | 8.513 | 500.9 | -61 | 2.46 | 5.25 |
| | 1.0 | 8.250 | 8.546 | 503.74 | -40 | 2.4 | 3.4 |
| | 1.5 | 8.277 | 8.597 | 510.01 | -30 | 2.39 | <2 |
| | 1.7 | 8.283 | 8.638 | 513.24 | -28 | 2.46 | <2 |
| | 2.0 | 8.280 | 8.650 | 513.95 | | | |
| y = | 0.3 | 8.217 | 8.379 | 489.92 | -90 | 2.38 | 6 |
| | 0.5 | 8.211 | 8.373 | 488.90 | -95 | 2.56 | 5.2 |
| | 1.0 | 8.182 | 8.188 | 474.69 | -96 | 2.33 | 3.4 |
| | 1.5 | 8.157 | 8.024 | 462.42 | -180 | 2.57 | <2 |
| | 1.7 | 8.138 | 7.916 | 454.00 | -188 | 2.54 | <2 |
| | 2.0 | 8.131 | 7.893 | 454.00 | | | |
| z = | 0.2 | 8.233 | 8.447 | 495.86 | -59 | 2.50 | 7.0 |
| | 0.4 | 8.230 | 8.464 | 496.44 | -57 | 2.51 | 7.0 |
| | 0.6 | 8.227 | 8.480 | 497.05 | -56 | 2.56 | 7.0 |
| | 0.8 | 8.228 | 8.494 | 497.92 | -45 | 2.53 | 7.0 |
| | 1.0 | 8.226 | 8.520 | 499.19 | -44 | 2.49 | 7.0 |

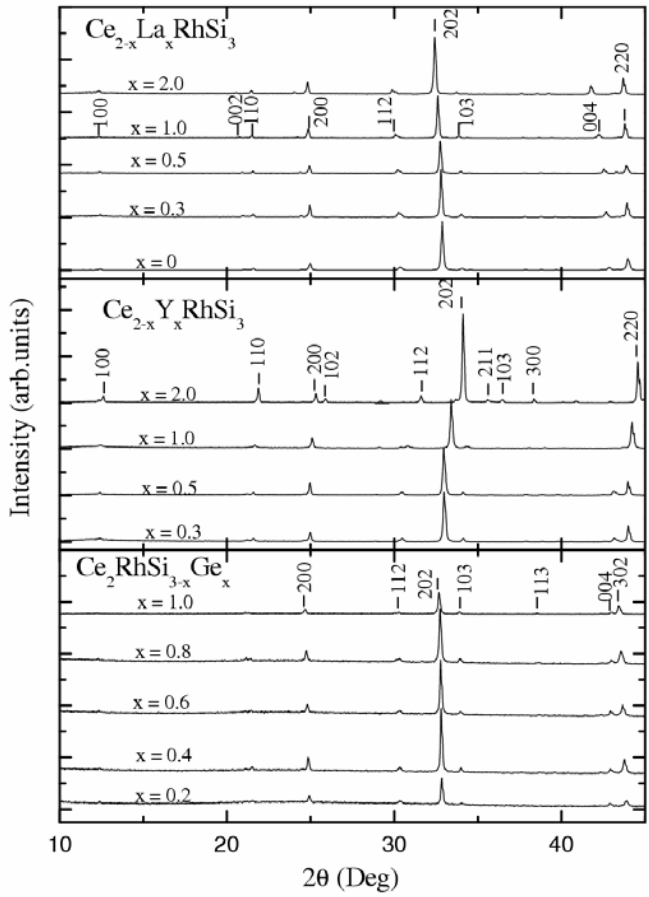

Figure 1: X-ray diffraction patterns (Cu K$_\alpha$) for Ce$_{2-x}$La$_x$RhSi$_3$, Ce$_{2-y}$Y$_y$RhSi$_3$, and Ce$_2$RhSi$_{3-z}$Ge$_z$,

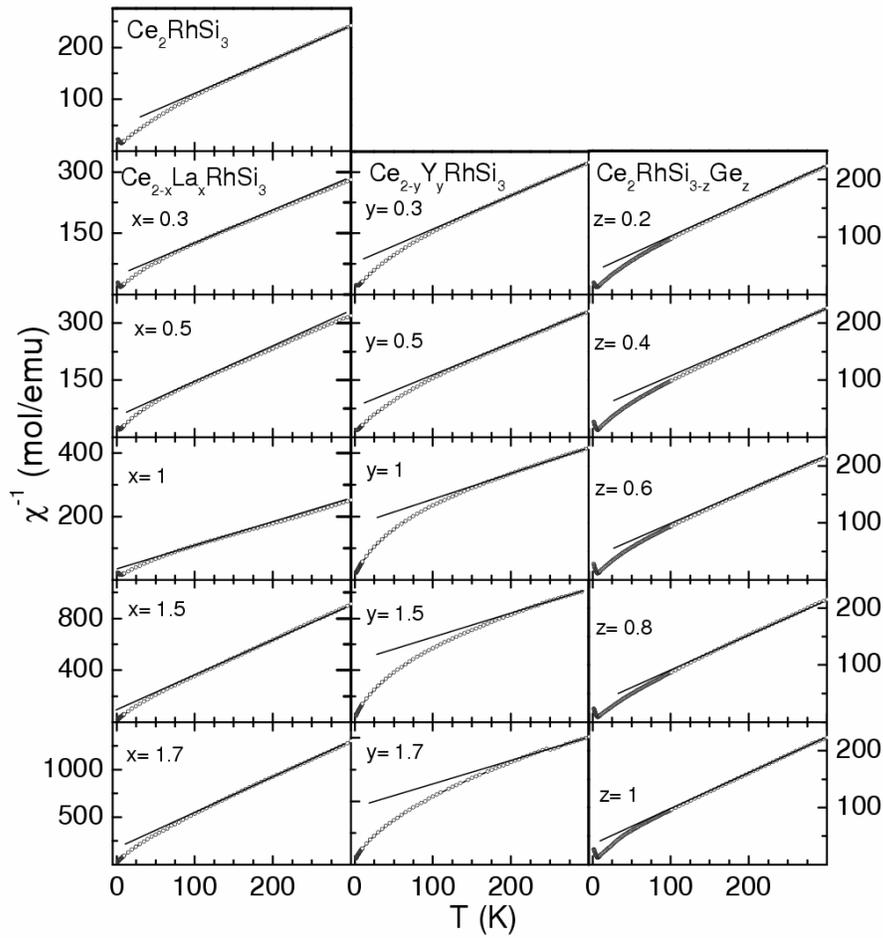

**Figure 2:** Inverse magnetic susceptibility as a function of temperature (1.8 – 300 K) for the alloys, $Ce_{2-x}La_xRhSi_3$, $Ce_{2-y}Y_yRhSi_3$, and $Ce_2RhSi_{3-z}Ge_z$. A line is drawn through high temperature linear region for each composition.

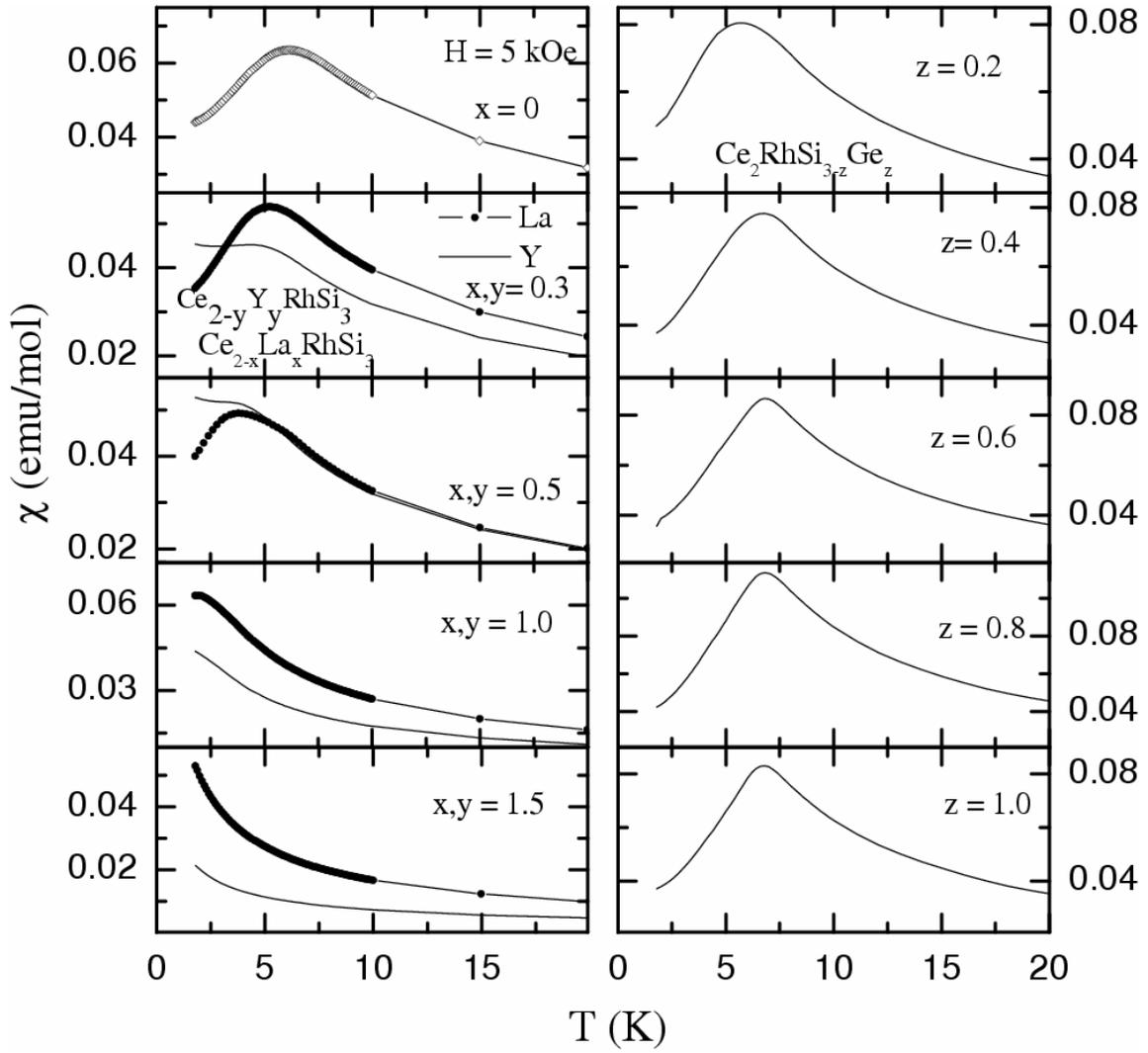

Figure 3: Magnetic susceptibility as a function of temperature in the low temperature range (to highlight the features due to magnetic ordering) for $Ce_{2-x}La_xRhSi_3$, $Ce_{2-y}Y_yRhSi_3$, and $Ce_2RhSi_{3-z}Ge_z$ alloys.

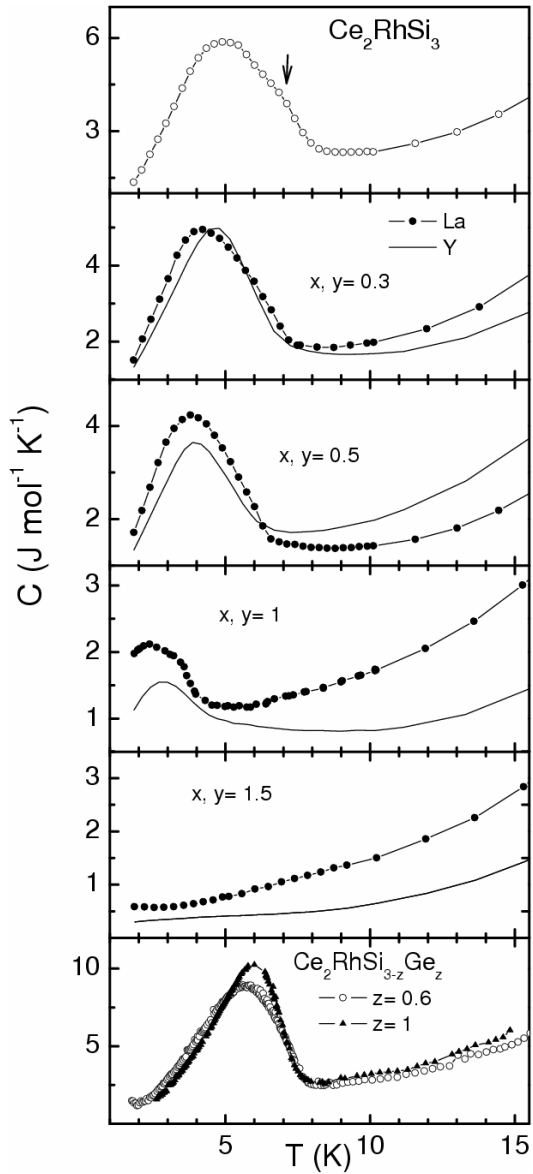

**Figure 4:** Heat capacity as a function of temperature for the alloys, $Ce_{2-x}La_xRhSi_3$, $Ce_{2-y}Y_yRhSi_3$, and $Ce_2RhSi_{3-z}Ge_z$ at low temperatures. The way the Néel temperature is typically obtained is marked by an arrow for x= 0.0.

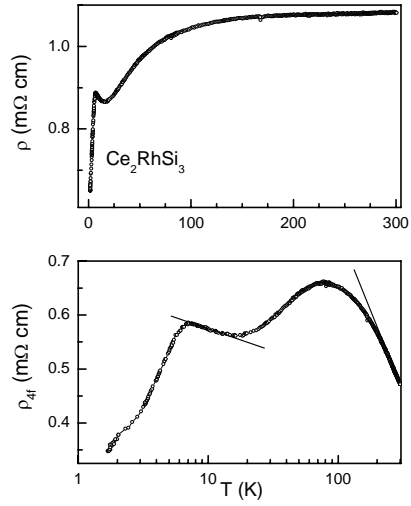

**Figure 5:** Temperature dependence of electrical resistivity (ρ) for $Ce_2RhSi_3$ (top) and the 4f contribution ($\rho_{4f}$) to ρ (bottom) for $Ce_2RhSi_3$ as described in the text. The two logarithmic slopes are also shown by lines in the bottom figure.

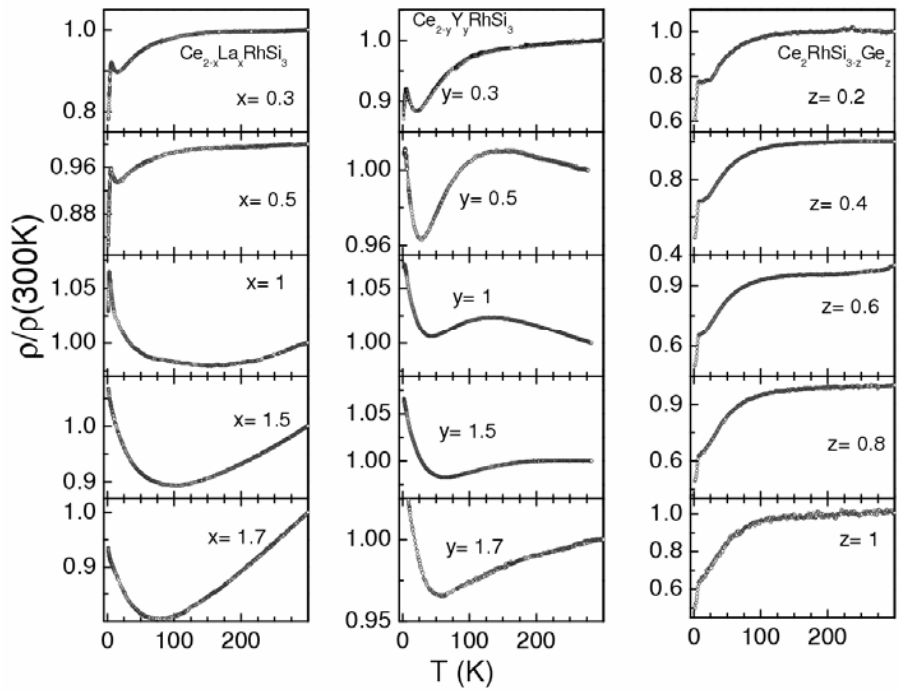

**Figure 6:** Temperature dependence (1.8-300 K) of electrical resistivity (ρ) for the alloys, $Ce_{2-x}La_xRhSi_3$, and $Ce_{2-y}Y_yRhSi_3$, and $Ce_2RhSi_{3-z}Ge_z$, normalized to respective 300 K values.

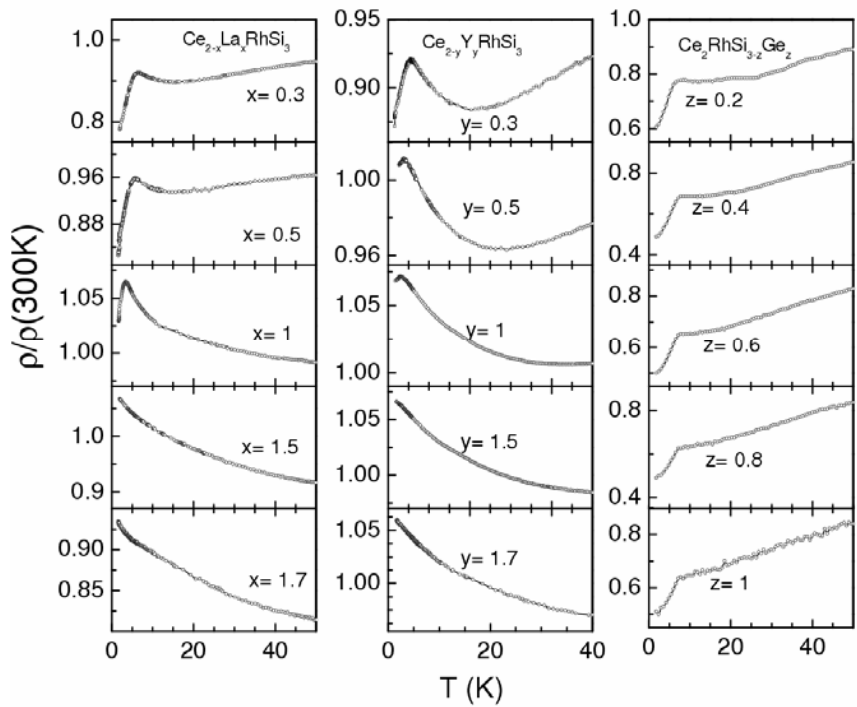

**Figure 7:** Low temperature (<50 K) behavior of electrical resistivity (ρ) for the alloys, $Ce_{2-x}La_xRhSi_3$, and $Ce_{2-y}Y_yRhSi_3$, and $Ce_2RhSi_{3-z}Ge_z$, normalized to respective 300 K values.

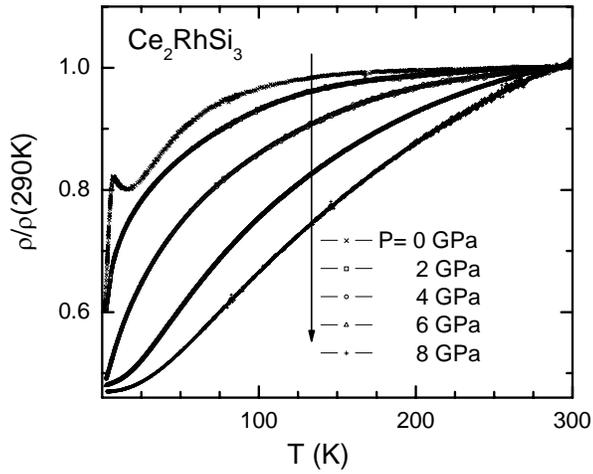

**Figure 8:** Temperature dependence of electrical resistivity (ρ) in the range 1.8-300 K for $Ce_2RhSi_3$ obtained under various external pressures employing cubic anvil pressure cell up to 8 GPa.

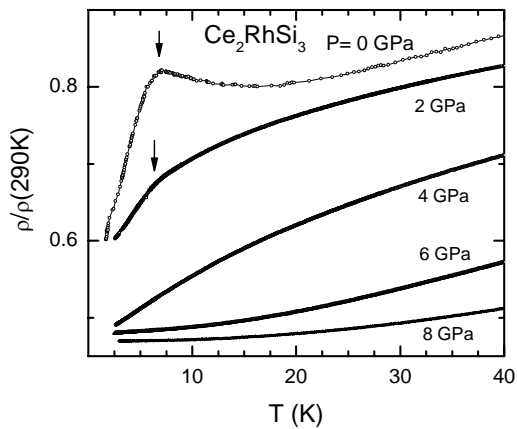

**Figure 9:** Temperature dependence of electrical resistivity (ρ) in the range 1.8-40 K for $Ce_2RhSi_3$ obtained under various external pressures employing cubic anvil pressure cell up to 8 GPa.

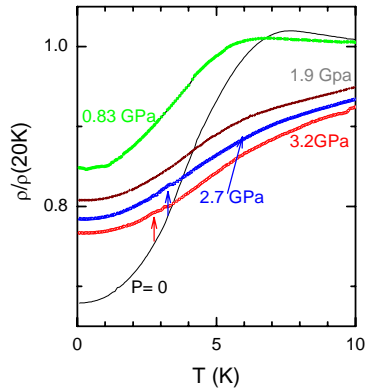

**Figure 10:** Temperature dependence of electrical resistivity (ρ) in the range 50 mK to 10 K for $Ce_2RhSi_3$ obtained under various external pressures employing a hybrid cylinder in the dilution refrigerator. Vertical arrows for P= 2.7 and 3.2 GPa data are marked where there is a weak step which can be attributed to $T_N$.

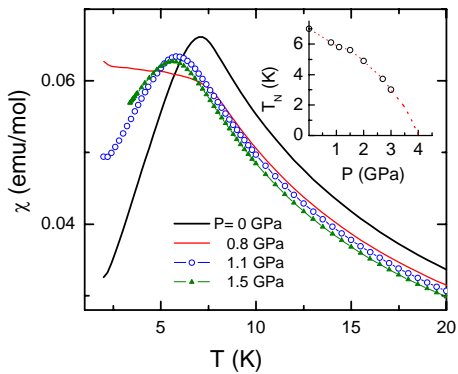

**Figure 11:** Magnetic susceptibility (χ) as a function of temperature for P= 0, 0.8, 1.1 and 1.5 GPa for $Ce_2RhSi_3$ measured employing a Cu-Be cell in a SQUID magnetometer in the presence of a magnetic field of 10 kOe. Inset shows a profile of $T_N$ as a function of pressure inferred from all the high pressure data.

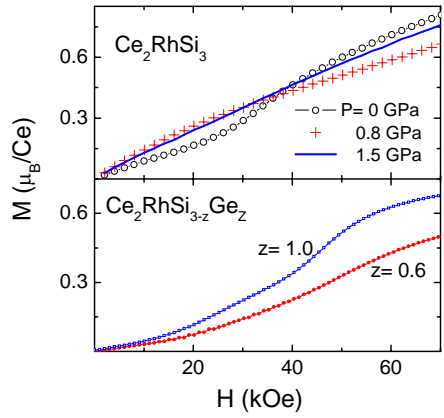

**Figure 12:** Isothermal magnetization behavior of $Ce_2RhSi_3$ under pressure as well as of Ge substituted alloys at 1.8 K.